\documentclass[epj,nopacs]{svjour}
\usepackage{amssymb}
\usepackage{amsmath}
\usepackage[dvips]{graphicx}
\usepackage{amsfonts}%
\providecommand{\U}[1]{\protect\rule{.1in}{.1in}}

\begin{document}

\title{The pseudoscalar hadronic channel contribution of the
light-by-light process to the muon (g-2)$_{\mu}$
within the nonlocal chiral quark model}
\titlerunning{The pseudoscalar hadronic channel contribution of the
light-by-light process ...}

\author{A.E. Dorokhov\inst{1} \and A.E. Radzhabov\inst{2} \and A.S. Zhevlakov\inst{2}
}                     
\institute{Bogoliubov Laboratory of Theoretical Physics, JINR, 141980 Dubna, Russia
\and Institute for System Dynamics and Control Theory, 664033 Irkutsk, Russia}
%

\abstract{
The light-by-light contribution to the anomalous magnetic moment of muon $(g-2)_\mu$
from the hadronic exchanges in the neutral pseudoscalar meson channel is calculated
in  the nonlocal chiral quark model. The full kinematic dependence of the
meson-two-photon vertices from the virtualities of the mesons and photons  is taken into
account. The status of various phenomenological and QCD short-distance constraints
is discussed and the comparison with the predictions of other models is performed.
It is demonstrated that the effect of the full kinematic dependence in the meson-photon vertices is to reduce the contribution of pseudoscalar
exchages to  $a_{\mu}^{\mathrm{PS,LbL}}$ by approximately factor 1.5 in comparison with the most of
previous estimates.
}

\maketitle

\section{Introduction}

The study of the anomalous magnetic moments (AMM) of leptons, $a=(g-2)/2$, has
played an important role in the development of the standard model. At the
present level of accuracy, the muon AMM, $a_{\mu},$ leads to the sensitivity
to effects of new physics, with particles in the mass range of a few hundred
GeV/c$^{2}$.

Comparison of the experimental measurements of the muon AMM
\cite{Bennett:2006fi} with the predictions of the standard model (see, for
example \cite{Jegerlehner:2011ti}) shows deviation by approximately 3$\sigma$.
The uncertainties of the standard model value for $a_{\mu}$ are dominated by
the uncertainties of the hadronic contributions $a_{\mu}^{\mathrm{Strong}},$
since their evaluation involves methods of nonperturbative QCD at large
distances. The contribution to the muon AMM to leading order in the QED
coupling constant $\alpha$ comes from the hadronic vacuum polarization and the
next-to-leading corrections consist of the contributions which are the
iteration of the leading order term plus the contribution from the
light-by-light (LbL) scattering process. The absolute values of the leading and
next-to-leading terms are differed by one order of magnitude, but the theoretical
accuracy of their extraction is comparable and dominates the overall
theoretical error of the standard model result. The hadronic vacuum
polarization contribution to the muon AMM is known with accuracy better than
one percent owing to the phenomenological analysis of the inclusive
$e^{+}e^{-}\rightarrow$ hadrons and $\tau\rightarrow$ hadrons data
\cite{Davier:2010nc}. This, however, is not the case for the LbL contribution
known from various QCD motivated approaches with accuracy of order $50\%$. It
is the LbL contribution, which will dominate the theoretical error of the
standard model prediction in the near future, and more realistic approaches
are needed for a better understanding.

In general, the LbL scattering amplitude is a complicated object for
calculations. It is a sum of different diagrams, the quark loop, the meson
exchanges, the meson loops and the iterations of these processes. Fortunately,
already in the first papers
\cite{Hayakawa:1995ps,Bijnens:1995cc,Hayakawa:1997rq} devoted to the
calculation of the hadronic LbL contribution, it was recognized that these
numerous terms show a hierarchy. This is related to existence of two small
parameters: the inverse number of colors $1/N_{c}$ and the ratio of the
characteristic internal momentum to the chiral symmetry parameter $m_{\mu
}/(4\pi f_{\pi})\sim0.1$. The latter suppresses the multiloop contributions,
so that the leading contributions are due to the quark loop diagram and the
two-loop diagrams with mesons in the intermediate state. Here the contribution
of the diagram with intermediate pion is enhanced by small pion mass in the
meson propagator.

Different approaches to the calculation of the pseudoscalar meson
contributions to the muon AMM from the light-by-light scattering process are
used. These approaches can be separated in two groups. The first one consists
of various versions of the vector meson dominance model (VMD)
\cite{Hayakawa:1997rq,Knecht:2001qf,Melnikov:2003xd,Nyffeler:2009tw,Cappiello:2010uy}
and the second one of QCD inspired effective models (EM). They include
different versions of the Nambu--Jona-Lasinio model
\cite{Bijnens:2001cq,Bartos:2001pg}, the models based on nonperturbative
quark-gluon dynamics, like the instanton liquid model 
\cite{Dorokhov:2008pw}, the Schwinger-Dyson model \cite{Fischer:2010iz,Goecke:2010if} (DSE)
or the holographic models (HM) \cite{Hong:2009zw}. To reduce the model
dependence of various approaches, different constraints on their parameter
space are employed. One kind of important constraints on the models follows
from the phenomenology of the two-photon widths of the pseudoscalar mesons
$\Gamma\left(  P\rightarrow\gamma\gamma\right)  $ and their transition form
factors $\mathrm{F}_{P\gamma\gamma^{\ast}}\left(  -M_{P}^{2};0,q^{2}\right)  $
first emphasized in \cite{Hayakawa:1997rq}. Another set of constraints follows
from the large momentum asymptotics for the meson transition form factors
\cite{Hayakawa:1997rq,Knecht:2001qf} and for the total light-by-light
scattering amplitude considered in \cite{Melnikov:2003xd,Dorokhov:2008pw},
obtained using perturbative QCD. In addition, the model amplitudes have to be
consistent with the 4-momentum conservation law. Finally, the model calculations
should be tested by calculations of the hadronic vacuum polarization to the
muon AMM
\cite{deRafael:1993za,Pallante:1994ee,Holdom:1993ad,Pivovarov:2001mw,Dorokhov:2004ze}.

The present work is devoted to the calculation of the pseudoscalar mesons
contribution of the hadronic light-by-light scattering process to the muon AMM
within the nonlocal chiral quark model (N$\chi$QM)
\cite{Anikin:2000rq,Dorokhov:2003kf,Scarpettini:2003fj}. In earlier work
\cite{Dorokhov:2008pw} the contribution of the pion exchange was estimated
within this approach in the chiral limit. Now we include the effect of the
current quark masses and extend the calculations also to $\eta$ and
$\eta^{\prime}$ meson exchanges. As was emphasized in \cite{Dorokhov:2008pw},
one of the main advantage of the effective model approaches like N$\chi$QM is
to control the full kinematical dependence of the LbL amplitude. The paper is
organized as follows. In Sect. 2 the basic elements of the effective model is
given. In Sect. 3 the properties of the three-point amplitude with arbitrary
pseudoscalar meson and photon virtualities are considered. The numerical
results on the light-by-light contribution in the pseudoscalar channel
obtained within the nonlocal chiral quark model are given in Sect. 4. Sect. 5
is devoted to discussion of the constraints for the LbL amplitude that allow
to diminish the model dependence of the estimates. Sect. 6 and 7 contain our
results in comparison with other model calculations and conclusions.

\section{N$\chi$QM Lagrangian, T matrix and $\eta-\eta^{\prime}$ mixing}

The Lagrangian of the $SU(3)\times SU(3)$ chiral quark model has the
form\footnote{For simplicity in this work we do not consider an extended model
that includes other structures besides the pseudoscalar (P) and scalar (S)
ones.}
\begin{align}
\mathcal{L}  &  =\bar{q}(x)(i\hat{\partial}-m_{c})q(x)+\frac{G}{2}[J_{S}%
^{a}(x)J_{S}^{a}(x)+J_{P}^{a}(x)J_{P}^{a}(x)]\nonumber\\
&  -\frac{H}{4}T_{abc}[J_{S}^{a}(x)J_{S}^{b}(x)J_{S}^{c}(x)-3J_{S}^{a}%
(x)J_{P}^{b}(x)J_{P}^{c}(x)], \label{33model}%
\end{align}
where $q\left(  x\right)  $ are the quark fields, $m_{c}$ is the diagonal
matrix of\ the quark current masses\footnote{We consider the isospin limit
$m_{c,u}=m_{c,d}\neq m_{c,s}$.}, $G$ and $H$ are the four- and six-quark
coupling constants. Second line in the Lagrangian represents the
Kobayashi--Maskawa--t`Hooft determinant vertex with the structural constant
\[
T_{abc}=\frac{1}{6}\epsilon_{ijk}\epsilon_{mnl}(\lambda_{a})_{im}(\lambda
_{b})_{jn}(\lambda_{c})_{kl},
\]
where $\lambda_{a}$ are the Gell-Mann matrices for $a=1,..,8$ and $\lambda
_{0}=\sqrt{2/3}I$. The nonlocal structure of the model is introduced via the
nonlocal quark currents
\begin{align}
J_{M}^{a}(x)=\int d^{4}x_{1}d^{4}x_{2}\,f(x_{1})f(x_{2})\, \bar{q}
(x-x_{1})\,\Gamma_{M}^{a}q(x+x_{2}),\nonumber
\end{align}
where $M=S,P$ and $\Gamma_{{S}}^{a}=\lambda^{a}$, $\Gamma_{{P}}=i\gamma^{5}\lambda^{a}$, and $f(x)$ is a form factor reflecting the nonlocal properties of the QCD\ vacuum.

The model (\ref{33model}) can be bosonized using the stationary phase
approximation which leads to the system of gap equations for the dynamical
quark masses $m_{d,i}\quad(i=u,d,s)$%
\begin{align}
m_{d,u}+GS_{u}+\frac{H}{2}S_{u}S_{s} &  =0,\nonumber\\
m_{d,s}+GS_{s}+\frac{H}{2}S_{u}^{2} &  =0,
\end{align}
with
\begin{align}
S_{i}=-8N_{c}\int\frac{d_{E}^{4}k}{(2\pi)^{4}}\frac{f^{2}(k^{2})m_{i}(k^{2}%
)}{D_{i}(k^{2})},
\end{align}
where $m_{i}(k^{2})=m_{c,i}+m_{d,i}f^{2}(k^{2})$, $D_{i}(k^{2})=k^{2}%
+m_{i}^{2}(k^{2})$ is the dynamical quark propagator obtained by solving the
Schwinger-Dyson equation, $f(k^{2})$ is the nonlocal form factor in the
momentum representation.

The vertex functions and the meson masses can be found from the Bethe-Salpeter
equation. For the separable interaction (\ref{33model}) the quark-antiquark
scattering matrix in pseudoscalar channel becomes
\begin{align}
&\mathbf{T}=\hat{\mathbf{T}}(p^{2})\delta^4\left(p_1+p_2-(p_3+p_4)\right) \prod \limits_{i=1}^4 f(p_i^2),\nonumber\\
&\hat{\mathbf{T}}(p^{2})=i\gamma_{5}\lambda_{k}\left(  \frac{1}{-\mathbf{G}%
^{-1}+\mathbf{\Pi}(p^{2})}\right)  _{kl}i\gamma_{5}\lambda_{l},
\end{align}
where $p_i$ are the momenta of external quark lines,
$\mathbf{G}$ and $\mathbf{\Pi}(p^{2})$ are the corresponding matrices of
the four-quark coupling constants and the polarization operators of
pseudoscalar mesons ($p=p_1+p_2=p_3+p_4$). The meson masses can be found from the zeros of
determinant $\mathrm{det}(\mathbf{G}^{-1}-\mathbf{\Pi}(-M^{2}))=0$ and the
$\hat{\mathbf{T}}$-matrix for the system of pseudoscalar mesons can be expressed in
the form
\begin{align}
\hat{\mathbf{T}}(p^{2})=\sum_{a=\pi^{0},\eta,\eta^{\prime}}\frac{\bar{V}_{a}(p^{2})\otimes
V_{a}(p^{2})}{-(p^{2}+M_{a}^{2})},\label{Tmatrix}%
\end{align}
where $M_{a}$ are the meson masses, $V_{a}(p^{2})$ are the vertex functions $\left(
\bar{V_{a}}(p^{2})=\gamma^{0}V_{a}^{\dag}(p^{2})\gamma^{0}\right)  $. In general case of
three unequal quark masses it is necessary to solve the $\pi^{0}-\eta
-\eta^{\prime}$ (or $\eta^{0}-\eta^{3}-\eta^{8}$) system factorizing in the
isospin limit into the $\pi^{0}$ and $\eta-\eta^{\prime}$ systems. For the
$\eta-\eta^{\prime}$ system it is convenient to diagonalize the scattering
matrix by orthogonal transformation
\begin{align}%
\begin{pmatrix}
\eta\\
\eta^{\prime}%
\end{pmatrix}
=%
\begin{pmatrix}
\cos\theta & -\sin\theta\\
\sin\theta & \cos\theta
\end{pmatrix}%
\begin{pmatrix}
\eta_{8}\\
\eta_{0}%
\end{pmatrix}
.
\end{align}
As a result, the mesonic vertex functions introduced in (\ref{Tmatrix})
become
\begin{align}
V_{\pi^{0}}\left(  p^{2}\right)   &  =i\gamma_{5}g_{\pi}(p^{2})\lambda
_{3} ,\\
V_{\eta}\left(  p^{2}\right)   &  =i\gamma_{5}g_{\eta}(p^{2})\left(
\lambda_{8}\cos\theta(p^{2})-\lambda_{0}\sin\theta(p^{2})\right),\nonumber\\
V_{\eta^{\prime}}\left(  p^{2}\right)   &  =i\gamma_{5}g_{\eta^{\prime}}%
(p^{2})\left(  \lambda_{8}\sin\theta(p^{2})+\lambda_{0}\cos\theta
(p^{2})\right) ,\nonumber
\end{align}
where $g_{a}(p^{2})$  and $\theta(p^{2})$ are the meson renormalization constants and mixing angles depending on the meson virtuality.
The renormalization constants are defined through the unrenormalized meson
propagators $D_{a}(p^{2})$ as
\begin{align}
{g_{a}^{2}}(p^{2})={-(p^{2}+M_{a}^{2})}D_{a}(p^{2}).
\end{align}
The $\mathbf{G}$-matrix elements have the form
\begin{align}
G_{00}&=G-\frac{H}{3}(2S_{u}+S_{s}),\nonumber\\
G_{88}&=G+\frac{H}{6}(4S_{u}-S_{s}),\\
G_{08}&=G_{80}=\frac{\sqrt{2}}{6}H(S_{u}-S_{s}),\nonumber\\
G_{\pi}&=G_{33}=G+\frac{H}{2}S_{s}.\nonumber
\end{align}
The matrix $\mathbf{\Pi}(p^{2})$ is diagonal in the quark-flavor basis and in
the singlet-triplet-octet basis it is given by
\begin{align}
\Pi_{00}(p^{2})&=\frac{1}{3}\left(  2\Pi_{uu}(p^{2})+\Pi_{ss}(p^{2})\right),\nonumber\\
\Pi_{88}(p^{2})&=\frac{1}{3}\left(  \Pi_{uu}(p^{2})+2\Pi_{ss}(p^{2})\right), \\
\Pi_{08}(p^{2})&=\Pi_{80}(p^{2})=\frac{\sqrt{2}}{3}\left(  \Pi_{uu}(p^{2})-\Pi_{ss}(p^{2})\right),\nonumber\\
\Pi_{\pi}(p^{2})&=\Pi_{33}(p^{2})=\Pi_{uu}(p^{2}),\nonumber
\end{align}
where the polarization operators are
\begin{align}
\Pi_{ij}(p^{2})&=8N_{c}\int\frac{d_{E}^{4}k}{(2\pi)^{4}}\frac{f^{2}(k_{+}^{2})f^{2}(k_{-}^{2}) }{D_{i}(k_{+}^{2})D_{j}(k_{-}^{2}) }\times\nonumber\\
&\times\left[  (k_{+}\cdot k_{-})+m_{i}(k_{+}^{2})m_{j}%
(k_{-}^{2})\right] ,
\end{align}
and $k_{\pm}=k\pm p/2$. The unrenormalized mesonic propagators for the
pseudoscalar mesons are
\begin{align}
D_{\pi}^{-1}(p^{2})&=-G_{\pi}^{-1}+\Pi_{\pi}(p^{2}),\nonumber\\
D_{\eta,\eta^{\prime}}^{-1}(p^{2})&=\frac{1}{2}\left[  (A+C)\mp \sqrt{(A-C)^{2}+4B^{2}}\right]  ,\nonumber\\
 A&=-G_{88}/\mathrm{det}(\mathbf{G})+\Pi_{00}(p^{2}),\\
 B&=+G_{08} /\mathrm{det}(\mathbf{G})+\Pi_{08}(p^{2}),\nonumber\\
 C&=-G_{00}/\mathrm{det}(\mathbf{G})+\Pi_{88}(p^{2}),\nonumber\\
&\mathrm{det}(\mathbf{G})=G_{00}G_{88}-G_{08}^{2}.\nonumber
\end{align}
The meson mixing angle depends on the virtuality
\begin{align}
\theta(p^{2})=\frac{1}{2}\arctan\left[  \frac{2B}%
{A-C}\right]  -\frac{\pi}{2}\Theta\left(  A-C\right).
\end{align}
Therefore $\theta_{\eta}=\theta(-M_{\eta}^{2})$ and $\theta
_{\eta^{\prime}}=\theta(-M_{\eta^{\prime}}^{2})$ are different for the
on-shell $\eta$ and $\eta^{\prime}$ mesons.

For numerical estimations we use the Gaussian nonlocal form factor
\begin{align}
f(k^{2})=\exp(-k^{2}/{2\Lambda^{2}}), \label{Gff}
\end{align}
and the model parameters obtained in \cite{Scarpettini:2003fj}. In that work
these parameters (the current quark masses $m_{c,i}$, the coupling constants
$G$ and $H$, and the nonlocality scale $\Lambda$) are fixed by requiring that
the model reproduces correctly the measured values of the pion and kaon
masses, the pion decay constant $f_{\pi}$, and the $\eta^{\prime}$ mass
(parameter sets $\mathrm{G}_{I}$, $\mathrm{G}_{IV}$) or the $\eta^{\prime
}\rightarrow\gamma\gamma$ decay constant $g_{\eta^{\prime}\gamma\gamma}$ (sets
$\mathrm{G}_{II}$, $\mathrm{G}_{III}$). For completeness, we present the
parameter sets $G_{I-IV}$ and the basic meson properties in Tables
\ref{table:ModParams} and \ref{table:MesonProp} as they were determined in
\cite{Scarpettini:2003fj}. The sets $\mathrm{G}_{I}$, $\mathrm{G}_{IV}$ vary
by different input for the nonstrange current quark mass, while $\mathrm{G}%
_{II}$, $\mathrm{G}_{III}$ are two solutions of the same fitting procedure.%

\begin{table*}[ht] \centering
\begin{tabular}
[c]{|c|c|c|c|c|c|c|c|}\hline
set & $m_{c,u}$ & $m_{d,u}$ & $m_{c,s}$ & $m_{d,s}$ & $\Lambda$ &
$G\Lambda^{2}$ & $H\Lambda^{5}$\\
& [MeV] & [MeV] & [MeV] & [MeV] & [MeV] &  & \\\hline
$\mathrm{G}_{I}$ & $8.5$ & $304.5$ & $223$ & $427$ & $709$ & $21.986$ &
$-1670.19$\\
$\mathrm{G}_{II}$ & $8.5$ & $304.5$ & $223$ & $439$ & $709$ & $22.898$ &
$-1557.28$\\
$\mathrm{G}_{III}$ & $8.5$ & $304.5$ & $223$ & $422$ & $709$ & $21.605$ &
$-1717.59$\\
$\mathrm{G}_{IV}$ & $7.5$ & $287.5$ & $199$ & $408$ & $768$ & $20.896$ &
$-1721.69$\\\hline
\end{tabular}
\caption{The model parameter sets obtained in
\cite{Scarpettini:2003fj}.}\label{table:ModParams}%
\end{table*}%

\begin{table*}[ht] \centering
\begin{tabular}
[c]{|c|c|c|c|c|c|c|}\hline
set & $M_{\pi}$ & $M_{\eta}$ & $M_{\eta^{\prime}}$ & $g_{\pi\gamma\gamma}$ &
$g_{\eta\gamma\gamma}$ & $g_{\eta^{\prime}\gamma\gamma}$\\
& [MeV] & [MeV] & [MeV] & [GeV$^{-1}$] & [GeV$^{-1}$] & [GeV$^{-1}$]\\\hline
$\mathrm{G}_{I}$ & $138.9$ & $516.5$ & $958.4$ & $0.2706$ & $0.3082$ &
$0.3752$\\
$\mathrm{G}_{II}$ & $138.9$ & $505.4$ & $878.6$ & $0.2706$ & $0.3259$ &
$0.3401$\\
$\mathrm{G}_{III}$ & $138.9$ & $520.7$ & $1006.4$ & $0.2706$ & $0.3011$ &
$0.3489$\\
$\mathrm{G}_{IV}$ & $139.0$ & $522.1$ & $>739.7$ & $0.2713$ & $0.3068$ &
\\\hline
exp & $134.9766$ & $547.8533$ & $957.78$ & $0.2744$ & $0.2726$ & $0.3423$\\
& $\pm0.0006$ & $\pm0.024$ & $\pm0.06$ & $_{-0.008}^{+0.009}$ & $\pm0.008$ &
$\pm0.014$\\\hline
\end{tabular}
\caption{The basic meson properties for different parametrizations obtained in
\cite{Scarpettini:2003fj}.}\label{table:MesonProp}%
\end{table*}%

\section{Anomalous triangles}

By using the nonlocal chiral quark model\footnote{For the pion, the
corresponding vertex in the chiral limit has been considered in
\cite{Dorokhov:2002iu}.}, the triangular diagram with external pseudoscalar
meson and two photon legs with arbitrary virtualities (Fig. \ref{fig:LbL}) can
be written as%
\begin{align}
&A\left(  \gamma^{\ast}\left(  q_{1},\epsilon_{1}\right)  \gamma^{\ast}\left(
q_{2},\epsilon_{2}\right)  \rightarrow P^{\ast}\left(  p\right)  \right)
=\\
&\quad
-ie^{2}\varepsilon_{\mu\nu\rho\sigma}\epsilon_{1}^{\mu}\epsilon_{2}^{\nu
}q_{1}^{\rho}q_{2}^{\sigma}\mathrm{F}_{P^{\ast}\gamma^{\ast}\gamma^{\ast}%
}\left(  p^{2};q_{1}^{2},q_{2}^{2}\right)  \nonumber,\label{ApiGG}%
\end{align}
with the photon momenta $q_{1,2}$ and the polarization vectors $\epsilon
_{1,2}$, $p=q_{1}+q_{2}$. For different pseudoscalar meson states one has%
\begin{align}
&\mathrm{F}_{\pi_{0}^{\ast}\gamma^{\ast}\gamma^{\ast}}\left(  p^{2};q_{1}^{2},q_{2}^{2}\right)  =g_{\pi}(p^{2})F_{u}\left(  p^{2};q_{1}^{2} ,q_{2}^{2}\right)  ,\nonumber\\
&\mathrm{F}_{\eta^{\ast}\gamma^{\ast}\gamma^{\ast}}\left(  p^{2};q_{1}^{2},q_{2}^{2}\right)     =\frac{g_{\eta}(p^{2})}{3\sqrt{3}}\times\nonumber\\
&\quad\times\biggl[  \left(5F_{u}\left(  p^{2};q_{1}^{2},q_{2}^{2}\right)  -2F_{s}\left(  p^{2};q_{1}^{2},q_{2}^{2}\right)  \right)  \cos\theta(p^{2})-  \nonumber\\
&\quad\, -\sqrt{2}\left(  5F_{u}\left(  p^{2};q_{1}^{2},q_{2}^{2}\right)  +F_{s}\left(  p^{2};q_{1}^{2},q_{2}^{2}\right)  \right)\sin\theta(p^{2})\biggr]  ,\nonumber\\
&\mathrm{F}_{\eta^{\prime\ast}\gamma^{\ast}\gamma^{\ast}}\left(  p^{2};q_{1}^{2},q_{2}^{2}\right)  =\frac{g_{\eta^{\prime}}(p^{2})}{3\sqrt{3}} \times\\
&\quad\times\biggl[ \left(  5F_{u}\left(  p^{2};q_{1}^{2},q_{2}^{2}\right)-2F_{s}\left(  p^{2};q_{1}^{2},q_{2}^{2}\right)  \right)  \sin\theta(p^{2})+  \nonumber\\
& \quad\, +\sqrt{2}\left(  5F_{u}\left(  p^{2};q_{1}^{2},q_{2}^{2}\right)  +F_{s}\left(  p^{2};q_{1}^{2},q_{2}^{2}\right)  \right) \cos\theta(p^{2}) \biggr] ,\nonumber
\end{align}
with

\begin{align}
&  F_{i}\left(  p^{2};q_{1}^{2},q_{2}^{2}\right)  =8\int\frac{d_{E}^{4} k}{(2\pi)^{4}}\frac{f(k_{1}^{2})f(k_{2}^{2})}{D_{i}(k_{1}^{2})D_{i}(k_{2} ^{2})D_{i}(k^{2})}\times\nonumber\\
&\, \times\left[
 m_{i}(k^{2})-\mathrm{m}_{i}^{(1)}(k_{1},k)J_{1}
 -\mathrm{m}_{i}^{(1)}(k_{2},k)J_{2}
   \right]  ,\nonumber \\%
&\quad J_{1}
=k^{2}+\frac{q_{2}^{2}(kq_{1})(k_{1}q_{1})-q_{1}^{2}(kq_{2})(k_{1}q_{2})}{q_{1}^{2}q_{2}^{2}-(q_{1}q_{2})^{2}},\label{Fi}\\ 
&\quad J_{2}\,
 =\,k^{2}+\frac{q_{1}^{2}%
(kq_{2})(k_{2}q_{2})-q_{2}^{2}(kq_{1})(k_{2}q_{1})}{q_{1}^{2}q_{2}^{2}%
-(q_{1}q_{2})^{2}},\nonumber
\end{align}
where $k_{1}=k+q_{1}$, $k_{2}=k-q_{2}$, $\mathrm{m}_{i}^{(1)}(k,p)=(m_{i}%
(k^{2})-m_{i}(p^{2}))/(k^{2}-p^{2})$ is the first order finite-difference
derivative. From (\ref{Fi}) one can easily obtain the expressions for some
special kinematics%
\begin{align}
&  F_{i}\left(  q_{1}^{2};q_{1}^{2},0\right)  =8\int\frac{d_{E}^{4}k}%
{(2\pi)^{4}}\frac{f(k_{1}^{2})f(k^{2})}{D_{i}(k_{1}^{2})D_{i}^{2}(k^{2}%
)}\times \label{Fi2}\\
& \quad\, \times\left[  m_{i}(k^{2})
-\mathrm{m}_{i}^{(1)}(k_{1},k)\overline{J}_{1}
-m_{i}^{\prime}(k^{2})\overline{J}_{2}
\right]  ,\nonumber\\
& \quad\quad \overline{J}_{1}\left(  k,q_{1}\right)  =(kq_{1})+\frac{2}{3}\left[k^{2}+2\frac{(kq_{1})^{2}}{q_{1}^{2}}\right]  ,\nonumber\\
&\quad\quad \overline{J}_{2}
  =\frac{4}{3}\left[  k^{2}-\frac{(kq_{1})^{2}}{q_{1}^{2}}\right]  ,\nonumber\\
&F_{i}\left(  0;0,0\right)  =\frac{1}{m_{d,i}}\biggl[ \frac{1}{4\pi^{2}}-\label{Fi3}\\
&\quad\quad-8m_{c,i}\int\frac{d_{E}^{4}k}{(2\pi)^{4}}\frac{m_{i}(k^{2})-2m_{i}^{\prime
}(k^{2})k^{2}}{D_{i}^{3}(k^{2})}\biggr]  ,\nonumber
\end{align}
In particular, the kinematics displayed in (\ref{Fi2}) is of special interest
for the hadronic exchange LbL calculations (see Eq. (\ref{amu}) below). In
(\ref{Fi3}) the first term is due to the axial anomaly, while the second term
represents the correction due to explicit breaking of the chiral symmetry by
current quark mass.

\section{Light-by-light hadronic contribution to the muon AMM in the
pseudoscalar meson channel}

\begin{figure}[ht]
\resizebox{0.45\textwidth}{!}{%
  \includegraphics{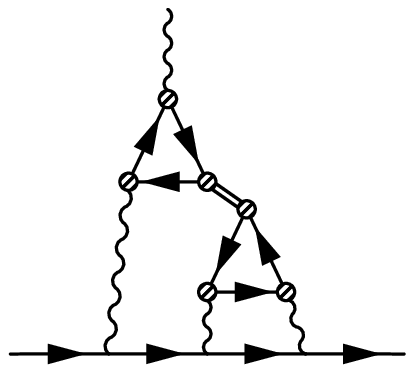}\includegraphics{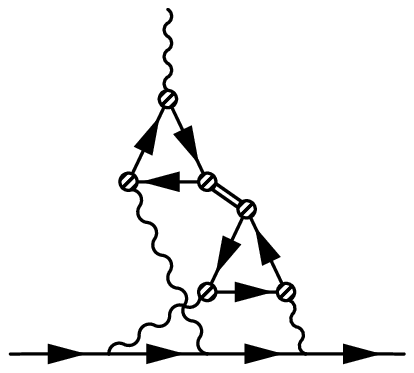}
}
\caption{Light-by-light contribution
from intermediate pseudoscalar meson exchanges.}%
\label{fig:LbL}%
\end{figure}

The light-by-light contribution due to exchanges of the light hadrons in the
intermediate pseudoscalar channel to the muon AMM is shown in Fig.
\ref{fig:LbL} and can be written in the form~\cite{Jegerlehner:2009ry}
\begin{align}
&  a_{\mu}^{\mathrm{LbL},\mathrm{PS}}=-\frac{2\alpha^{3}}{3\pi^{2}}%
\int\limits_{0}^{\infty}dq_{1}^{2}\int\limits_{0}^{\infty}dq_{2}^{2}%
\int\limits_{-1}^{1}dt\sqrt{1-t^{2}}\frac{1}{q_{3}^{2}}\times\nonumber \\
&  \times\sum_{a=\pi^{0},\eta,\eta^{\prime}}\biggl[  2\frac{\mathrm{F}%
_{a^{\ast}\gamma^{\ast}\gamma^{\ast}}\left(  q_{2}^{2};q_{1}^{2},q_{3}%
^{2}\right)  \mathrm{F}_{a^{\ast}\gamma^{\ast}\gamma}\left(  q_{2}^{2}%
;q_{2}^{2},0\right)}{q_{2}^{2}+M_{a}^{2}}I_{1} 
\nonumber\\
&\quad +\frac{\mathrm{F}_{a^{\ast}\gamma^{\ast}\gamma^{\ast}}\left(
q_{3}^{2};q_{1}^{2},q_{2}^{2}\right)  \mathrm{F}_{a^{\ast}\gamma^{\ast}\gamma
}\left(  q_{3}^{2};q_{3}^{2},0\right)  }{q_{3}^{2}+M_{a}^{2}}I_{2}
\biggr] \label{amu} ,
\end{align}
where $q_{3}=-\left(  q_{1}+q_{2}\right)  .$ The functions $I_{1}$ 
and $I_{2}$ 
are

\begin{align}
&I_{1}
  =q_{1}^{2}\Biggl[  \left\langle \frac
{1}{D_{1}}\right\rangle \left(  \frac{(q_{1}q_{2})}{2}-q_{2}^{2}\left( 1-t^{2}\right)  \right)  +\left\langle \frac{pq_{2}}{D_{1}}\right\rangle+\nonumber\\
&\,+\left\langle \frac{1}{D_{1}\cdot D_{2}}\right\rangle q_{2}^{2}\left( 1-t^{2}\right)  \left(  q_{2}^{2}-2M_{\mu}^{2}\right)  \Biggr]  -\frac{(q_{1}q_{2})}{2}\,, \label{I1}
\end{align}
\begin{align}
&I_{2}
  =2q_{2}^{2}\Biggl[  \left\langle \frac{1}{D_{2}}\right\rangle\left(  q_{1}^{2}+(q_{1}q_{2})\right)-\left\langle \frac{pq_{1}}{D_{2}}\right\rangle -\nonumber\\
& -\left\langle \frac{1}{D_{1}\cdot D_{2}}\right\rangle q_{1}^{2}\left(  q_{1}^{2}+(q_{1}q_{2})+M_{\mu}^{2}\left(  1-t^{2}\right)  \right)  \Biggr]+ \nonumber\\
&  +\left\langle \frac{1}{D_{1}}\right\rangle q_{1}^{2}(q_{1}q_{2}%
)-(q_{1}q_{2}), \label{I2}%
\end{align}
are obtained \cite{Jegerlehner:2009ry} after averaging over the directions of
muon momentum $p$%
\begin{align}
\left\langle ...\right\rangle =\frac{1}{2\pi^{2}}\int d\Omega\left(
\widehat{p}\right)  ...
\end{align}
In (\ref{I1}) and
(\ref{I2}) the notations are ($D_{1}=\left(p+q_{1}\right)^{2}+M_{\mu}^{2}$, $D_{2}=\left(p-q_{2}\right)^{2}+M_{\mu}^{2}$)
\begin{align}
& \left\langle \frac{1}{D_{1}}\right\rangle =\frac{R_{1}-1}{2M_{\mu}^{2}}\,,\quad
 \left\langle \frac{1}{D_{2}}\right\rangle =\frac{R_{2}-1}{2M_{\mu }^{2}}\,,\nonumber\\
& \left\langle \frac{1}{D_{1}\cdot D_{2}}\right\rangle =\frac{1}{M_{\mu}^{2}\left\vert q_{1}\right\vert \left\vert q_{2}\right\vert x}\arctan\left[  \frac{zx}{1-zt}\right]  \,,\nonumber\\
&  \left\langle \frac{pq_{1}}{D_{2}}\right\rangle =-(q_{1}q_{2})\frac{\left( 1-R_{2}\right)  ^{2}}{8M_{\mu}^{2}}\,\,,\,\,\,\label{Not}\\
& \left\langle \frac{pq_{2}} {D_{1}}\right\rangle =(q_{1}q_{2})\frac{\left(  1-R_{1}\right)  ^{2}}{8M_{\mu}^{2}}\,\,,\nonumber\\
& \quad
t=\frac{(q_{1}q_{2})}{\left\vert q_{1}\right\vert \left\vert q_{2}\right\vert },
\quad x=\sqrt{1-t^{2}}\, ,\,\,R_{i}=\sqrt
{1+\frac{4M_{\mu}^{2}}{q_{i}^{2}}}\,,\nonumber\\
&\quad z=\frac{q_{1}q_{2}}{4M_{\mu}^{2}
}\left(  1-R_{1}\right)  \left(  1-R_{2}\right)  \,,\,\,\,\nonumber
\end{align}
and $M_{\mu}$ is the muon mass $\left(  p^{2}=-M_{\mu}^{2}\right)  $.

\begin{figure}[t]
\resizebox{0.45\textwidth}{!}{%
  \includegraphics{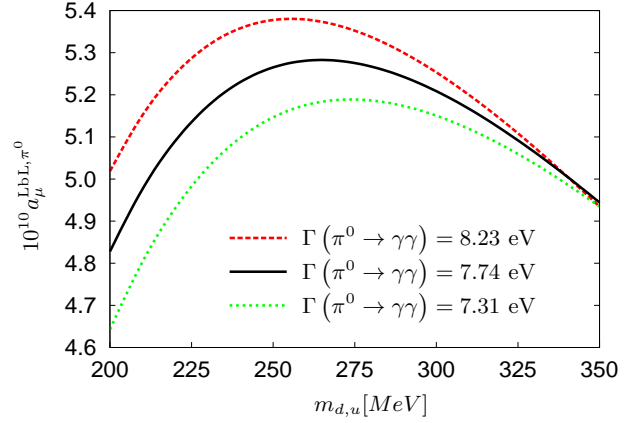}
}
\caption{Light-by-light contribution to
the muon AMM from the neutral pion exchange as a function of the dynamical
quark mass. The band between dotted lines corresponds to the error interval
for the pion two-photon width.}%
\label{fig:PionLbL}%
\end{figure}

It is instructive to investigate the pion contribution to the muon AMM
$a_{\mu}^{\mathrm{LbL,\pi^{0}}}$ for the $SU(2)$ reduction of the nonlocal
model (\ref{33model}). In this case, in the isospin limit, there are three
model parameters: the current $m_{c,u}$ and dynamical $m_{d,u}$ quark masses
and the nonlocality parameter $\Lambda$. Varying one parameter, say $m_{d,u}$,
in some region, one can fix other parameters by using as input the pion mass
and the two-photon decay constant of the neutral pion\footnote{The properties
of the charged pion, namely the mass and the weak decay constant, are used as
a rule for fixing the model parameters. In general, the difference between
fixing the parameter space for the charged pion or for the neutral pion is due
to small isotopic invariance breaking corrections induced by the
electro-magnetic interaction and inequality of the nonstrange quark masses. We
believe that for the problem under consideration it is better to fix the model
parameters by the neutral pion properties.}. The pion mass is known with high
accuracy, but the two-photon decay constant has an experimental uncertainty at
the level of 3\%. We investigate the dependence of $a_{\mu}^{\mathrm{LbL,\pi
^{0}}}$ on this uncertainty. Namely, the dynamical quark mass is taken in the
typical interval $200$--$350$ MeV and then other parameters are fitted by the
pion mass and the two-photon decay constant in correspondence with the pion
lifetime given within the error range in \cite{Nakamura:2010zzi}. The
dependency of $a_{\mu}^{\mathrm{LbL,\pi^{0}}}$ as a function of $m_{d,u}$
within the error interval for the two-photon decay constant is shown in Fig.
\ref{fig:PionLbL} and, thus, we get our conservative estimate $a_{\mu}^{\mathrm{LbL,\pi
^{0}}}=(5.01\pm0.37)\cdot10^{-10}$.

The results for the nonlocal $SU(3)$ model (\ref{33model}) with the parameter
sets $\mathrm{G}_{I-IV}$ are given in Table \ref{table:1}. Note, that the
results for $a_{\mu}^{\mathrm{LbL,\pi^{0}}}$ are in the error range consistent
with the results obtained in the nonlocal $SU(2)$ model and that $a_{\mu
}^{\mathrm{LbL,\eta}}$ and $a_{\mu}^{\mathrm{LbL,\eta^{\prime}}}$ are one
order less than the contribution to the muon AMM due to the pion.%

\begin{table}[ht] \centering
\begin{tabular}
[c]{|c|c|c|c|c|c|}\hline
set & $\pi^{0}$ & $\eta$ & $\eta^{\prime}$ & $\eta+\eta^{\prime}$ & $\pi
^{0}+\eta+\eta^{\prime}$\\\hline
$\mathrm{G}_{I}$  & $5.05$ & $0.55$ & $0.27$ & $0.82$ & $5.87$\\
$\mathrm{G}_{II}$ & $5.05$ & $0.59$ & $0.48$ & $1.08$ & $6.13$\\
$\mathrm{G}_{III}$& $5.05$ & $0.53$ & $0.18$ & $0.71$ & $5.76$\\
$\mathrm{G}_{IV}$ & $5.10$ & $0.49$ & $0.25$ & $0.74$ & $5.84$\\\hline
\end{tabular}
\caption{The contribution of pseudoscalar mesons to the muon AMM
$a_{\protect\mu }^{\mathrm{LbL}}$ obtained within N$\chi$QM for different
parametrizations. All numbers are given in $10^{-10}$.}\label{table:1}%
\end{table}%

\section{Constraints on $P\gamma\gamma$ vertex and LbL amplitude}

A first phenomenological constraint on the anomalous vertex $\mathrm{F}_{P^{\ast}%
\gamma^{\ast}\gamma^{\ast}}\left(  q_{3}^{2};q_{1}^{2},q_{2}^{2}\right)  $ is
put when both photons and the meson are on-shell. The vertex $\mathrm{F}%
_{P\gamma\gamma}\left(  -M_{P}^{2};0,0\right)  $ is normalized by the
experimentally measured two-photon decay widths
\begin{align}
\mathrm{F}_{P\gamma\gamma}\left(  -M_{P}^{2};0,0\right)  \equiv g_{P\gamma
\gamma}=\sqrt{\frac{64\pi\Gamma\left(  P\rightarrow\gamma\gamma\right)
}{\left(  4\pi\alpha\right)  ^{2}M_{P}^{3}}}. \nonumber 
\end{align}

\begin{figure}[tb]
\resizebox{0.45\textwidth}{!}{\includegraphics{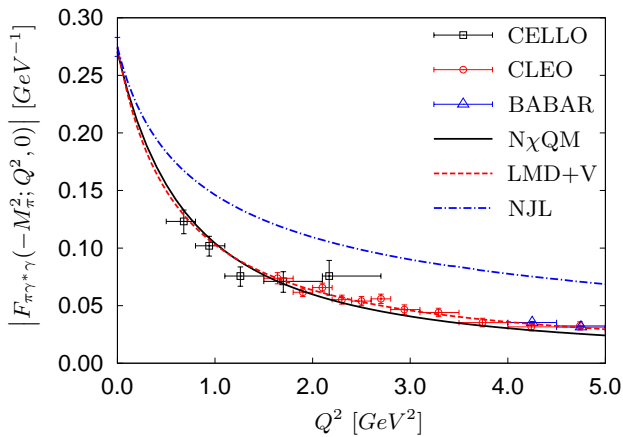}}
\caption{The $\gamma^{\ast}\gamma\rightarrow\pi^{0}$ transition form factor in N$\chi$QM (this work,
parameter set $\mathrm{G}_{I}$ is used), LMD+V \cite{Knecht:2001qf} and NJL
\cite{Bartos:2001pg} models in comparison with the CELLO\cite{Behrend:1990sr},
CLEO\cite{Gronberg:1997fj} and BABAR \cite{Aubert:2009mc} data.}%
\label{fig:PionFF}%
\end{figure}

\begin{figure}[tb]
\resizebox{0.45\textwidth}{!}{\includegraphics{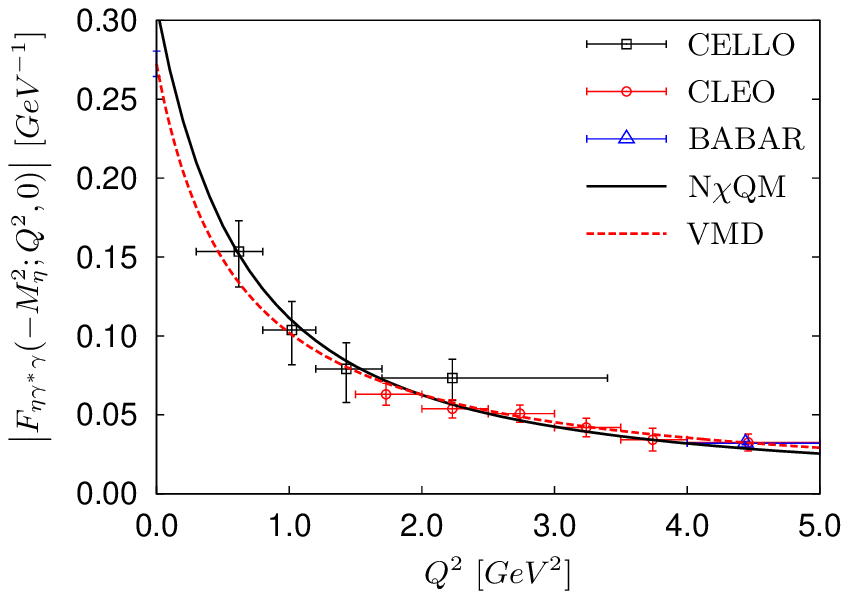}}
\caption{The $\gamma^{\ast} \gamma\rightarrow\eta$ transition form factor in N$\chi$QM (this work,
$\mathrm{G}_{I}$ set) and VMD \cite{Knecht:2001qf} models in comparison with
the CELLO\cite{Behrend:1990sr}, CLEO\cite{Gronberg:1997fj} and BABAR
\cite{:2011hk} data.}%
\label{fig:EtaFF}%
\end{figure}

\begin{figure}[tb]
\resizebox{0.45\textwidth}{!}{\includegraphics{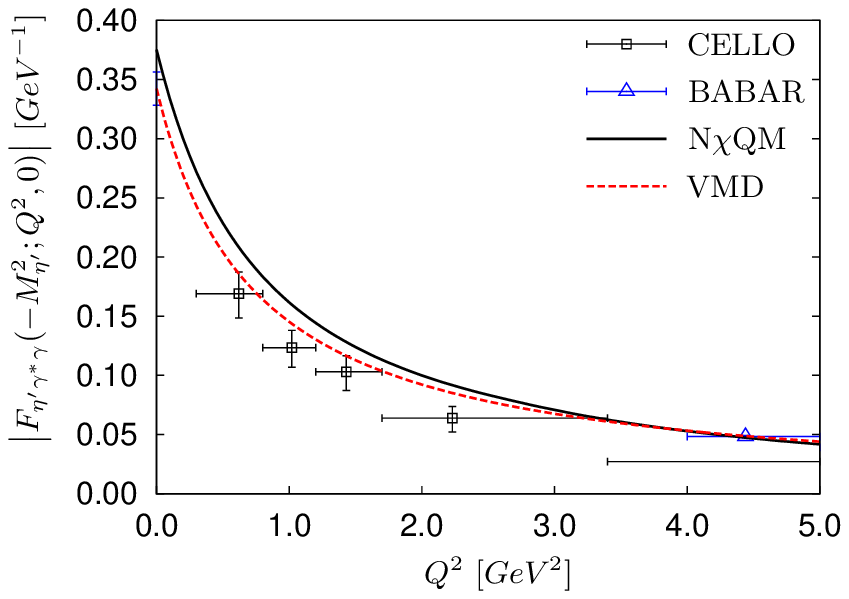}}
\caption{The $\gamma^{\ast} \gamma\rightarrow\eta^{\prime}$ transition form factor in N$\chi$QM (this
work, $\mathrm{G}_{I}$ set) and VMD \cite{Knecht:2001qf} models in comparison
with the CELLO \cite{Behrend:1990sr} and BABAR \cite{:2011hk} data points.}%
\label{fig:EtaPrFF}%
\end{figure}

A second phenomenological constraint is, that the vertex for special
kinematics, when the meson and one of the photon are on-shell, $\mathrm{F}%
_{P\gamma\gamma^{\ast}}\left(  -M_{P}^{2};0,q^{2}\right)  $, has to fit the
data on the pseudoscalar meson transition form factors available from the
measurements of the CELLO \cite{Behrend:1990sr}, CLEO \cite{Gronberg:1997fj}
and BABAR \cite{Aubert:2009mc,:2011hk} collaborations\footnote{As was
discussed in \cite{Nyffeler:2009uw,Dorokhov:2009dg,Dorokhov:2009jd}, new, very interesting
data of the BABAR collaboration \cite{Aubert:2009mc,:2011hk} at high momentum
transfer of the photon do not affect very much the low energy observables.}.
The transition form-factors for $\pi$, $\eta$, $\eta^{\prime}$ mesons
evaluated in the N$\chi$QM for the $\mathrm{G}_{I}$ parameter set are shown in
Figs. \ref{fig:PionFF},\ref{fig:EtaFF},\ref{fig:EtaPrFF}. One can see the
reasonable agreement of the N$\chi$QM (this work) and LMD+V
(VMD)\footnote{Some details about the LMD+V and VMD\ models are given in Sect.
6.} \cite{Knecht:2001qf} models predictions with existing experimental data,
while the NJL model with the parameter set taken in \cite{Bartos:2001pg} is
not perfect in explanation of the data.

A first QCD constraint \cite{Brodsky:1981rp} is, that at large photon
virtualities $q_{1}^{2}+q_{2}^{2}\rightarrow\infty$ the meson transition form
factor has the asymptotic behavior as
\begin{align}
\mathrm{F}_{P\gamma^{\ast}\gamma^{\ast
}}\left(  -M_{P}^{2};q_{1}^{2},q_{2}^{2}\right)  \sim1/\left(  q_{1}^{2}%
+q_{2}^{2}\right)   \nonumber
\end{align}
and for symmetric kinematics it fixes the asymptotic
coefficient. In the case of pion one has%
\begin{align}
&\mathrm{F}_{\pi\gamma^{\ast}\gamma^{\ast}}\left(  -M_{\pi}^{2};q^{2}%
,q^{2}\right)  \approx\frac{1}{q^{2}}\frac{2f_{\pi}}{3}\Biggl(  1-\frac{5}{6}\frac{\alpha_{s}\left(  q^{2}\right)  }{\pi}+\nonumber\\
&\quad\quad\quad\quad+O\left(  \alpha_{s}^{2}\right)  +O\left(  q^{-2}\right)  \Biggr)  . \label{BLsymLimit}%
\end{align}
Note, that even, if both in the N$\chi$QM and the LMD+V approaches the
$1/q^{2}$ behavior is satisfied and power corrections are taken into account,
these models are still incomplete to describe the radiative corrections
$O\left(  \alpha_{s}\right)  $ calculated within the perturbative QCD
\cite{Novikov:1983jt}.

A second QCD constraint is to match the total light-by-light amplitude with
three virtual $\left(q_{i},i=1,2,3\right)$ and one external magnetic
$\left(q_{4}\rightarrow0\right)$ field
\begin{align}
\mathcal{M}\left(q_{1}^{2},q_{2}^{2},q_{3}^{2}\right)  \equiv\alpha^{2}N_{c}\mathrm{Tr}\left[  \widehat{Q}%
^{4}\right]  \mathcal{A}\left(  q_{1}^{2},q_{2}^{2},q_{3}^{2}\right)  \nonumber
\end{align}
to the perturbative QCD asymptotics for the configurations when some of photon
virtualities are large \cite{Melnikov:2003xd}. There are only two distinct
kinematic regimes in the light-by-light scattering amplitudes: the Euclidean
momenta of the three photons are comparable in magnitude $q_{1}^{2}\approx
q_{2}^{2}\approx q_{3}^{2}\gg \Lambda_{\mathrm{QCD}}^{2},$ or one of the momenta
is much smaller than the other two $q_{1}^{2}\approx q_{2}^{2}\gg q_{3}%
^{2}\gg \Lambda_{\mathrm{QCD}}^{2}$. The second limit was analyzed in
\cite{Melnikov:2003xd} using the OPE with the result that the amplitude is
factorized in the short-distance factor $\sim1/q_{1}^{2}$ times the
nondiagonal correlator $\left(  VA\widetilde{V}\right)  $ between the axial
and electromagnetic current in the external magnetic field%
\begin{align}
&\mathcal{A}_{PS}\left(  q_{1}^{2},q_{2}^{2},q_{3}^{2}\right)  \overset
{q_{1}^{2}\approx q_{2}^{2}\gg q_{3}^{2}}{=}\label{APS}\\
&\quad=\sum_{a=3,8,0}W^{\left(  a\right)
}G_{2}^{(a)}\left(  q_{3}^{2},q_{2}^{2},q_{1}^{2}\right)  \left\{
f_{2}\overline{f}_{1}\right\}  \left\{  \overline{f}f_{3}\right\}
,\nonumber%
\end{align}
where%
\begin{align*}
&  G_{2}^{(a)}\left(  q_{3}^{2},q_{2}^{2},q_{1}^{2}\right)  \overset{q_{1}%
^{2}\approx q_{2}^{2}\gg q_{3}^{2}\gg \Lambda_{\mathrm{QCD}}^{2}}{=}\frac{2}%
{q_{1}^{2}}w_{L}^{\left(  a\right)  }\left(  q_{3}^{2}\right)  ,\\
&  W^{\left(  3\right)  }=\frac{1}{4},\quad W^{\left(  8\right)  }=\frac
{1}{12},\quad W^{\left(  0\right)  }=\frac{2}{3},
\end{align*}
$w_{L}^{\left(  a\right)  }\left(  q_{3}^{2}\right)  $ is the longitudinal
part of the $VA\widetilde{V}$ correlator with the corresponding flavor
structure of the axial current, $f_{i}^{\mu\nu}=q_{i}^{\mu}\epsilon_{i}^{\nu
}-q_{i}^{\nu}\epsilon_{i}^{\mu}$ are the field strength tensors, the braces
denote traces of the products of the matrices $f_{i}^{\mu\nu}$. In
(\ref{APS}), the amplitude $\mathcal{A}_{PS}$ is the part of the total
amplitude that is relevant to hadronic exchange in the pseudoscalar channel.
The properties of the $VA\widetilde{V}$ correlator was studied in detail in
\cite{Vainshtein:2002nv} where it was shown that in the chiral limit of
massless current quarks, due to properties of the axial anomaly, in the
perturbative QCD one has for all $a$%
\begin{align}
w_{L}^{\left(  a\right)  }\left(  q^{2}\right)  =\frac{2}{q^{2}},\label{wLT}%
\end{align}
moreover, these expressions for $w_{L}^{\left(  3,8\right)  }\left(
q^{2}\right)  $\ are exact QCD results for nonsinglet axial currents valid at
any $q^{2}$. \ This result was also confirmed within the effective instanton
liquid model (N$\chi$QM) \cite{Dorokhov:2005pg}. In\ \cite{Dorokhov:2005hw}
it was also demonstrated how the anomalous longitudinal part of the correlator
$w_{L}^{\left(  0\right)  }\left(  q^{2}\right)  $ is modified at low momenta
$\left(  q^{2}\lesssim m_{\eta^{\prime}}^{2}\right)  $ due to the presence of
the $U_{A}\left(  1\right)  $ anomaly. Thus within OPE one has, independently
of $a$, the QCD\ asymptotic constraint%
\begin{align}
G_{2}\left(  q_{3}^{2},q_{2}^{2},q_{1}^{2}\right)  \overset{q_{1}^{2}\approx
q_{2}^{2}\gg q_{3}^{2}\gg \Lambda_{\mathrm{QCD}}^{2}}{=}\frac{4}{q_{1}^{2}%
q_{3}^{2}}.\label{G2ope}%
\end{align}

The function $G_{2}\left(  q_{3}^{2},q_{2}^{2},q_{1}^{2}\right)  $ was given
in \cite{Melnikov:2003xd} as a result of explicit calculations in perturbative
theory for the light-by-light scattering amplitude where the photon-photon
interaction is mediated by the loop of massless quarks. This function is
symmetric under the permutation of the last two arguments and in specific
kinematics it is reduced to%
\begin{align}
&  G_{2}\left(  s,s,s\right)  =\frac{8}{3s^{2}},\label{G2as1}\\
&  G_{2}\left(  s_{3},s,s\right)  =\frac{8}{s_{3}\left(  s_{3}-4s\right)
^{2}}\biggl[  \left(  2s+s_{3}\right)  \ln\frac{s_{3}}{s}+\label{G2as2}\\
&\quad +4s-s_{3}+2s\left(
s-s_{3}\right)  J\left(  s_{3},s,s\right)  \biggr]  ,\nonumber\\
&  G_{2}\left(  s,s,s_{1}\right)  =\frac{4}{ss_{1}\left(  s_{1}-4s\right)
^{2}}\biggl[  \left(  4s^{2}-s_{1}^{2}\right)  \ln\frac{s}{s_{1}}+\label{G2as3}\\
&\quad+2s\left(
4s-s_{1}\right)  -2s\left(  s_{1}^{2}+2s^{2}-3ss_{1}\right)  J\left(
s,s,s_{1}\right)  \biggr] \nonumber ,%
\end{align}
where $s_{i}=q_{i}^{2}$. The triangle function $J\left(  s_{3},s_{2}%
,s_{1}\right)  $ (see, for example, \cite{Davydychev:1995mq}) is symmetric in
all its arguments and for the kinematics considered it is equal to
\begin{align}
&J\left(  s_{3},s,s\right)  =\label{Jtriangl}\\
&=\left\{
\begin{array}
[c]{l}%
\frac{1}{s}\frac{2}{\sqrt{x\left(  1-x/4\right)  }}\mathrm{Cl}_{2}\left(
\arccos\left(  1-\frac{x}{2}\right)  \right), \, \text{if }4s>s_{3},\\
\frac{1}{s_{3}}\frac{1}{\lambda}\left[  2\ln^{2}\left(  \frac{1-\lambda}%
{2}\right)  -4\mathrm{Li}_{2}\left(  \frac{1-\lambda}{2}\right)  -\ln
^{2}\left(  y\right)  +\frac{\pi^{2}}{3}\right],  \\
\quad\quad\quad\quad\quad\quad\quad\quad\quad\quad \text{ if } 4s<s_{3},
\end{array}
\right.  \nonumber %
\end{align}
where $x=\frac{s_{3}}{s}$ $\left(  0\leq x\leq4\right)  ,$ $y=\frac{s}{s_{3}}$
$\left(  0\leq y\leq1/4\right)  ,$ $\lambda=\sqrt{1-4y},$ $\mathrm{Cl}%
_{2}\left(  \theta\right)  $ is the Clausen integral function%
\[
\mathrm{Cl}_{2}\left(  \theta\right)  =-\int_{0}^{\theta}dt\ln\left\vert
2\sin\frac{t}{2}\right\vert ,
\]
and $\mathrm{Li}_{2}\left(  \eta\right)  $ is the dilogarithm (Spence)
function%
\[
\mathrm{Li}_{2}\left(  \eta\right)  =-\int_{0}^{1}dt\frac{\ln\left(  1-\eta
t\right)  }{t}.
\]
Asymptotically, when some photon virtualities are large, one has
\begin{align}
&  G_{2}\left(  s_{3},s,s\right)  \overset{s\gg s_{3}}{=}\frac{1}{ss_{3}}\Biggl[
4+\frac{s_{3}}{s}\left(  \frac{4}{3}\ln\left(  \frac{s_{3}}{s}\right)
-\frac{2}{9}\right)+\nonumber  \\
&\quad\quad\quad+O\left(  \left(  \frac{s_{3}}{s}\right)  ^{2}\right)
\Biggr]  ,\label{G21}\\
&  G_{2}\left(  s,s,s_{1}\right)  \overset{s\gg s_{1}}{=}\frac{4}{s^{2}}\Biggl[
-\frac{1}{3}\ln\left(  \frac{s_{1}}{s}\right)  +\frac{5}{9}+O\left(
\frac{s_{1}}{s}\right)  \Biggr]  ,\label{G22}%
\end{align}
and
\begin{align}
&  G_{2}\left(  s_{3},s,s\right)  \overset{s\ll s_{3}}{=}-\frac{8}{s_{3}^{2}%
}\Biggl[  \ln\left(  \frac{s}{s_{3}}\right)  +1+\nonumber\\
&+\frac{s}{s_{3}}\left(  2\ln
^{2}\left(  \frac{s}{s_{3}}\right)  +10\ln\left(  \frac{s}{s_{3}}\right)
+4+\frac{2}{3}\pi^{2}\right)  +\nonumber\\
&\quad\quad\quad\quad+O\left(  \left(  \frac{s}{s_{3}}\right)
^{2}\right)  \Biggr]  ,\label{G23}\\
&  G_{2}\left(  s,s,s_{1}\right)  \overset{s\ll s_{1}}{=}-\frac{4}{ss_{1}%
}\Biggl[  \ln\left(  \frac{s}{s_{1}}\right)+\nonumber\\
&  +\frac{s}{s_{1}}\left(  2\ln
^{2}\left(  \frac{s}{s_{1}}\right)  +8\ln\left(  \frac{s}{s_{1}}\right)
+2+\frac{2}{3}\pi^{2}\right) +\nonumber\\
&\quad\quad\quad\quad +O\left(  \left(  \frac{s}{s_{1}}\right)
^{2}\right)  \Biggr]  .\label{G24}%
\end{align}
These asymptotics have to play important role in constraining the effective
nonperturbative models when calculating the loops of dynamical quarks
\cite{Dorokhov:2008pw}.

In some recent works, however, it was claimed and used a so-called "new QCD
constraint" on the pion exchange LbL contribution to the muon $g-2$
\cite{Nyffeler:2009tw,Hong:2009zw,Cappiello:2010uy}. This "constraint" is an
attempt to restrict the behavior of the pion-two-photon vertex when the pion
is off-shell. In particular, in \cite{Nyffeler:2009tw} it was claimed that it
would be followed from QCD that
\begin{align}
\mathrm{F}_{\pi^{\ast}\gamma^{\ast}\gamma}^{\mathrm{oLMDV}}\left(  q^{2}%
;q^{2},0\right)  \overset{q^{2}\gg \Lambda_{\mathrm{QCD}}^{2}}{=}\frac{1}%
{3}f_{0}\chi, \label{OffNyffeler}%
\end{align}
where $f_{0}$ is the pion-decay constant in the chiral limit and $\chi$ is the
quark condensate magnetic susceptibility in the presence of a constant
external electromagnetic field. Let us show that (\ref{OffNyffeler}) is
inconsistent with OPE and QCD constraints (\ref{G2as1})-(\ref{G22}).

Indeed, in \cite{Nyffeler:2009tw} the off-shell pion form factor is
constructed as (off-shell LMD + V model)%
\begin{align}
&\mathrm{F}_{\pi^{\ast}\gamma^{\ast}\gamma^{\ast}}^{\mathrm{oLMDV}}\left(
q_{3}^{2};q_{1}^{2},q_{2}^{2}\right)  =\label{VMDL}\\
&\, =\frac{f_{\pi}}{3}\frac{-q_{1}^{2}q_{2}^{2}\left( q_{1}^{2}+q_{2}^{2}+q_{3}^{2}\right)  +P_{H}^{V}\left(
q_{3}^{2};q_{1}^{2},q_{2}^{2}\right)}{\left( q_{1}^{2}+M_{1}^{2}\right)
\left(  q_{1}^{2}+M_{2}^{2}\right)  \left(  q_{2}^{2}+M_{1}^{2}\right)
\left(  q_{2}^{2}+M_{2}^{2}\right)  },\nonumber\\
&\,P_{H}^{V}\left(  q_{3}^{2};q_{1}^{2},q_{2}^{2}\right)     =h_{1}\left(
q_{1}^{2}+q_{2}^{2}\right)  ^{2}+h_{2}q_{1}^{2}q_{2}^{2}+\nonumber\\
&\,+h_{3}\left(
q_{1}^{2}+q_{2}^{2}\right)  q_{3}^{2}  +h_{4}q_{3}^{4}-h_{5}\left(  q_{1}^{2}+q_{2}^{2}\right)  -h_{6}q_{3}%
^{2}+h_{7},\nonumber
\end{align}
where $M_{(1,2)}$ are the the masses of the lowest vector meson states, and
the coefficients in (\ref{VMDL}) are fixed from the above mentioned
constraints. Specifically, \ $h_{7}$ is due to the pion two-photon decay
width, $h_{1}=0$ by requirement (\ref{BLsymLimit}), $h_{5}$ is from the fit of
the CLEO data \cite{Gronberg:1997fj}, $h_{2}$ is related to higher power
corrections in (\ref{BLsymLimit}) and finally $\left(  h_{3}+h_{4}\right)
M_{1}^{-2}M_{2}^{-2}=\chi$ is fixed from the "new constraint"
(\ref{OffNyffeler}). As a result from (\ref{VMDL}) one get the asymptotics of
the pion-exchange part of the total LbL amplitude as%
\begin{align}
&{\normalsize G}_{2,PS}^{\mathrm{oLMDV}}\left(  q_{3}^{2},q_{2}^{2},q_{1}%
^{2}\right)     =\\
&\quad\quad=\frac{\mathrm{F}_{\pi^{\ast}\gamma^{\ast}\gamma^{\ast}%
}\left(  q_{3}^{2};q_{1}^{2},q_{2}^{2}\right)  \mathrm{F}_{\pi^{\ast}%
\gamma^{\ast}\gamma}\left(  q_{3}^{2};q_{3}^{2},0\right)  }{q_{3}^{2}+M_{\pi
}^{2}},\nonumber\\
&{\normalsize G}_{2,PS}^{\mathrm{oLMDV}}\left(s,s,s\right) \overset{s\gg \Lambda_{\mathrm{QCD}}^{2}}%
{=}-\frac{1}{3}f_{0}^{2}\chi\frac{1}{s^{2}},\label{AsympLMDV}\\
&{\normalsize G}_{2,PS}^{\mathrm{oLMDV}}\left(s_{3},s,s\right) \overset{s\gg s_{3}\gg \Lambda_{\mathrm{QCD}}^{2}}%
{=}-\frac{2}{9}f_{0}^{2}\chi\frac{1}{s s_{3}},\label{AsympLMDV2}%
\end{align}
Comparing the asymptotic coefficients in (\ref{AsympLMDV}) and
(\ref{AsympLMDV2}) with corresponding QCD coefficients in (\ref{G2as1}) and
(\ref{G21}) it is clear that the results based on the "constraint"
(\ref{OffNyffeler}) are in contradiction with the QCD asymptotics obtained in
a model independent way. We have already commented in \cite{Dorokhov:2008pw}
on the origin of this confusion as an incorrect identification of the
$\left\langle PVV\right\rangle $ correlator with $\left\langle \pi
VV\right\rangle .$ In the latter case, it appears the physical hadronic
pseudoscalar current that leads to much higher suppression (\ref{OffNLQM})
than it is shown in (\ref{OffNyffeler}) where the local pseudoscalar current
is used. The behavior (\ref{OffNyffeler}) is in variance with the prediction
of the N$\chi$QM where from (\ref{Fi2}) one has at large $q^{2}$
\begin{align}
\mathrm{F}_{P^{\ast}\gamma^{\ast}\gamma}^{\mathrm{N\chi QM}}\left(
q^{2};q^{2},0\right)  \overset{q^{2}\gg \Lambda_{\mathrm{QCD}}^{2}}{\sim}%
\exp\left(  -q^{2}/\Lambda^{2}\right), \label{OffNLQM}%
\end{align}
that does not violate the OPE results.

At this point, for completeness, we remind that the LMD+V model
\cite{Knecht:2001qf} corresponds to the expression (\ref{VMDL}) at $q_{3}%
^{2}=0$. The VMD model, used in particular in \cite{Knecht:2001qf} to describe
the form factors of the $\eta$ and $\eta^{\prime}$ mesons, is defined as%
\begin{align}
\mathrm{F}_{P\gamma^{\ast}\gamma^{\ast}}^{\mathrm{VMD}}\left(  0;q_{1}%
^{2},q_{2}^{2}\right)  =\frac{g_{P\gamma\gamma} \Lambda_{M}^{4}}{\left(
q_{1}^{2}+\Lambda_{M}^{2}\right)  \left(  q_{2}^{2}+\Lambda_{M}^{2}\right)  },
\label{Fvmd}%
\end{align}
where the parameters $\Lambda_{M}$ are taken from the fit of CLEO data
\cite{Gronberg:1997fj}: $\Lambda_{\pi}=776\pm22$ MeV,
$\Lambda_{\eta}=774\pm29$ MeV, $\Lambda_{\eta^{\prime}}=859\pm28$ MeV.

As a final remark we note, that from the results of
\cite{Vainshtein:2002nv,Melnikov:2003xd,Dorokhov:2005pg} it is easy to get the
correct expression that includes the magnetic susceptibility $\chi$ in the
asymptotic expansion of the total light-by-light amplitude. Indeed, by using
the OPE as it was shown in \cite{Vainshtein:2002nv} (see also
\cite{Dorokhov:2005pg}) due to spontaneous breaking of the chiral symmetry
there is the power correction to $w_{L}\left(  q^{2}\right)  $ that is linear
in the current quark mass%
\begin{align}
\Delta w_{L}\left(  q^{2}\right)  =\frac{16\pi^{2}}{3}\frac{m_{c}\left\vert
\left\langle \overline{q}q\right\rangle \right\vert \chi}{q^{4}}. \label{dwL}%
\end{align}
Then, we have the power correction to the total light-by-light amplitude
\begin{align}
\Delta G_{2}\left( q_{3}^{2}, q_{1}^{2},q_{2}^{2}\right)  \overset{q_{1}^{2}\approx q_{2}^{2}\gg q_{3}^{2}\gg \Lambda_{\mathrm{QCD}}^{2}}{=}\frac
{32\pi^{2}}{3}\frac{m_{c}\left\vert \left\langle \overline{q}q\right\rangle
\right\vert \chi}{q_{1}^{2}q_{3}^{4}}, \label{dG2}%
\end{align}
which is suppressed by the current quark mass and by extra power of large
momentum squared.

Thus our conclusion is that the effective models, in particular the instanton
liquid based N$\chi$QM model, are able to satisfy to most of phenomenological
and QCD\ constraints and take into account the full kinematic dependence on
virtualities of the mesons and photons. As to VMD based models, it is the
LMD+V model \cite{Knecht:2001qf} for the pion form factor that satisfies to
many constraints. The most serious problem of the VMD models is that they do
not take into account the dependence of the meson-two-photon vertex on the
meson virtuality and thus the momentum conservation is violated in this
vertex. We discuss this point with more details in the next section.

\section{Results and comparison with other models}

The results of different approaches in calculation of the pseudoscalar meson
contributions to the muon AMM from light-by-light scattering mechanism are
given in Table \ref{table:2}. For the pion contribution to the muon AMM in
N$\chi$QM we take the numbers from the $SU(2)$ version of the model. For
$\eta$ and $\eta^{\prime}$ we use the average of the results obtained with
different parameter sets, see Table \ref{table:1}.%

\begin{table*}[ht] \centering
\begin{tabular}
[c]{|l|l|l|l|l|}\hline
Model & $\pi^{0}$ & $\eta$ & $\eta^{\prime}$ & $\pi^{0}+\eta+\eta^{\prime}%
$\\\hline
VMD \cite{Hayakawa:1997rq} & $5.74$ & $1.34$ & $1.19$ & $8.27(0.64)$\\
ENJL \cite{Bijnens:2001cq} & $5.6$ &  &  & $8.5(1.3)$\\
LMD+V, VMD \cite{Knecht:2001qf} & $5.8(1.0)$ & $1.3(0.1)$ & $1.2(0.1)$ &
$8.3(1.2)$\\
NJL \cite{Bartos:2001pg} & $8.18(1.65)$ & $0.56(0.13)$ & $0.80(0.17)$ &
$9.55(1.66)$\\
(LMD+V)$^{\prime},$VMD\cite{Melnikov:2003xd} & $7.97$ & $1.8$ & $1.8$ &
$11.6(1.0)$\\
N$\chi$QM \cite{Dorokhov:2008pw} & $6.5(0.2)$ &  &  & \\
HM \cite{Hong:2009zw} & $6.9$ & $2.7$ & $1.1$ & $10.7$\\
DIP, VMD \cite{Cappiello:2010uy} & $6.54(0.25)$ &  &  & \\
DSE \cite{Goecke:2010if} & $5.75(0.69)$ & $1.36(0.30)$ & $0.96(0.21)$ &
$8.07(1.20)$\\
This work (N$\chi$QM) & $5.01(0.37)$ & $0.54$ & $0.30$ & $5.85$\\\hline
\end{tabular}

\caption{The light-by-light contribution to the muon AMM from the hadronic exchanges
in the neutral pseudoscalar channel $a_{\protect\mu }^{\mathrm{LbL}}$ obtained in
different works. All numbers are given in $10^{-10}$.
Note, that as it was checked in \cite{Bijnens:2007pz,Dorokhov:2008pw}, the (LMD+L)'
model \cite{Melnikov:2003xd} numerically predicts $7.97$ instead of $7.65$
quoted originally in \cite{Melnikov:2003xd}.
}\label{table:2}%
\end{table*}%

\begin{figure}[h]
\resizebox{0.45\textwidth}{!}{\includegraphics{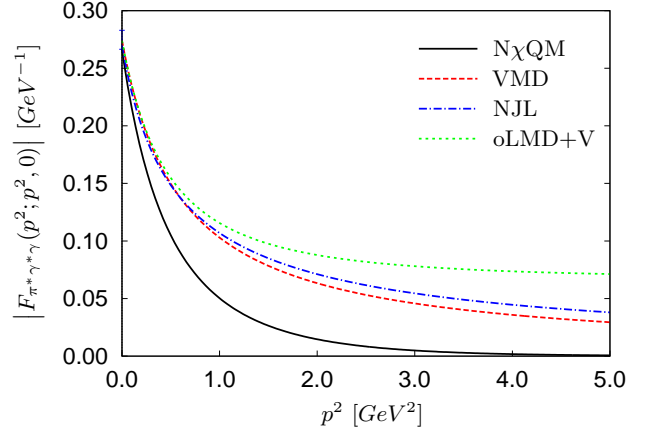}}
\caption{The photon-pion vertex for the special kinematics when the pion and one photon are off-shell
and have the same momenta, and another photon is real. The N$\chi$QM, Eq.
(\ref{Fi2}), is given by the solid line, the VMD, Eq. (\ref{Fvmd}), is by the
dashed line, the NJL result \cite{Bartos:2001pg} is by dash-dotted line, and
oLMD+V, Eq. (\ref{VMDL}), is by the short dashed line.}%
\label{SpecKin}%
\end{figure}

One can conclude that within the N$\chi$QM the pseudoscalar meson
contributions to muon AMM are systematically lower then the results obtained
in the other works. This effect can be understood considering more carefully
the off-shell meson-two-photon vertex $F_{P^{\ast}\gamma^{\ast}\gamma}
(p^{2};p^{2},0)$ entering the light-by-light contribution Eq. (\ref{amu}). The
comparison of the N$\chi$QM predictions for the pion in this kinematics
(\ref{Fi2}) with other models is presented in Fig. \ref{SpecKin}. One can see
that the N$\chi$QM leads to stronger suppression of the vertex at all momenta.
This explains why the results for $a_{\mu}^{\mathrm{LbL,\pi^{0}}}$ obtained
within the N$\chi$QM is smaller comparing with other calculations.

\begin{figure}[h]
\resizebox{0.45\textwidth}{!}{\includegraphics{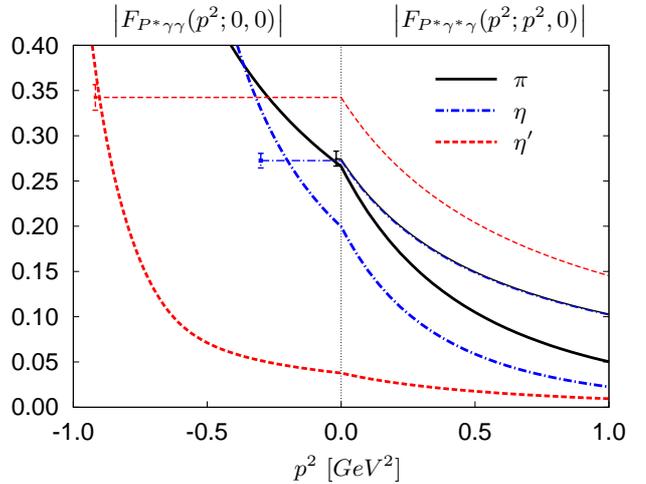}}
\caption{Plots of the $\pi^{0},\eta$ and $\eta^{\prime}$ vertices $F_{P^{\ast}\gamma\gamma}(p^{2};0,0)$ in the timelike
region and $F_{P^{\ast}\gamma^{\ast}\gamma}(p^{2};p^{2},0)$ in the spacelike
region in N$\chi$QM model (thick lines) and VMD, Eq. (\ref{Fvmd}), model (thin
lines). The points with error bars correspond to the physical points of the
meson decays into two photons. The VMD curves for $\pi^{0}$ and $\eta$ are
almost indistinguishable.}%
\label{fig:CompFF}%
\end{figure}

The situation with $\eta$`s is more dramatic. For illustration in Fig.
\ref{fig:CompFF} we present on the same plot the vertex $F_{P^{\ast}%
\gamma\gamma}(p^{2};0,0)$ in the timelike region $p^{2}\leq0$ and the vertex
$F_{P^{\ast}\gamma^{\ast}\gamma}(p^{2};p^{2},0)$ in the spacelike region
$p^{2}\geq0$ as they look in the N$\chi$QM and VMD models. These two special
kinematics match at zero virtuality $p^{2}=0$. The remarkable feature of this
construction is that the first kinematics is connected with the decay of
pseudoscalar mesons into two photons at physical points $F_{P\gamma\gamma
}(-M_{M}^{2};0,0)=g_{P\gamma\gamma}$, while the second kinematics is relevant
for the light-by-light contribution to the muon AMM (\ref{amu}). Thus, the
part of Fig. \ref{fig:CompFF} at $p^{2}<0$ describes the transition of the
pion-two-photon vertex from the physical points of meson masses to the point
with zero virtuality, which is the edge point of the interval where the
integrand of (\ref{amu}) is defined. In VMD type of models, including LMD+V
model, there is no such dependence on the meson virtuality. Thus, the value of
this vertex at zero meson virtuality is the same as the value of the vertex at
the physical points of meson masses, $F_{P\gamma\gamma}^{\mathrm{VMD}}%
(p^{2}=-M_{\eta,\eta^{\prime}}^{2};0,0)=F_{P\gamma\gamma}^{\mathrm{VMD}%
}(0;0,0)$. However, the $\eta$ and $\eta^{\prime}$ mesons are much heavier
than the pion and such extrapolation is too crude. One can see that for $\eta$
and particularly for $\eta^{\prime}$ the difference between the values of the
vertex at physical and zero virtuality points is large in the N$\chi$QM,
$F_{P\gamma\gamma}(p^{2}=-M_{\eta,\eta^{\prime}}^{2};0,0)\gg$ $F_{P\gamma
\gamma}(0;0,0)$. Thus, the contributions of the $\eta$ and $\eta^{\prime}$
mesons to the muon AMM evaluated in N$\chi$QM are strongly suppressed as
compared with the VMD\ results that can only be considered as upper estimates
of these contributions\footnote{Recently another attempt to take into account
the full kinematic dependence of the pion-photon vertex is given in the DSE
approach \cite{Goecke:2010if}. In this paper, the meson's contribution is
investigated using the pion-pole approximation with an off-shell prescription.
Although this prescription leads to a small suppression a full T-matrix calculation
therein is highly desirable. We thank Richard Williams for clarifying this point.}.

\section{Conclusions}

In this paper we have analyzed the pseudoscalar meson exchange contributions
of the light-by-light process to $a_{\mu}^{\mathrm{PS,LbL}}$
in the nonlocal chiral quark model. The basic new elements of our work are the
inclusion of the full kinematic dependence of the pseudoscalar meson-photon
vertex on virtualities of photons and mesons and the effect of current quark
masses. This study extends the previous work \cite{Dorokhov:2008pw} where the
full kinematic dependence was considered for the pion in the chiral limit. The
dependence of the vertices on the pion virtuality diminishes the result by
about 20-30\% as compared to the case where this dependence is neglected. We
find that the most significant effect occurs for the contributions of the
$\eta$ and $\eta^{\prime}$ mesons to the muon AMM. In this case the results
are reduced by factor about 3 in comparison with the results obtained in other
effective quark models where the kinematic dependence was neglected. Thus our
main conclusion is that within the realistic nonlocal chiral quark model the
total contribution of pseudoscalar exchanges to $a_{\mu}^{\mathrm{PS,LbL}}$ is by
approximately factor 1.5 less than the most of previous estimates.

We reviewed the phenomenological and QCD constraints on the hadronic
light-by-light scattering amplitude. It is shown that the effective models
like N$\chi$QM are able to fit most of these constraints, while the models
based on the vector meson dominance violate the momentum conservation law and
lead to overestimation of the meson exchanges contribution to the muon AMM. We
have also demonstrated that one of the constraints for the LbL amplitude
recently discussed in the literature \cite{Nyffeler:2009tw} is not justified.

An important next step in the investigations of the light-by-light scattering
within the nonlocal chiral quark model is to perform an extension to the
so-called complete calculation (see
\cite{Bijnens:1995xf,Hayakawa:1997rq,Bartos:2001pg}) that includes the scalar
and axial-vector meson exchanges, as well as takes into account the quark and
meson box diagrams. Due to contact terms arising in the nonlocal model, a
calculation of the box diagrams is technically rather involved and will be
presented elsewhere.

\begin{acknowledgement}
This work is supported in part by the program Development of Scientific
Potential in Higher Schools (2.2.1.1/1483, 2.1.1/1539), the Federal Target
Program "Research and Training Specialists in Innovative Russia 2009-2013"
(02. 740.11.5154), JINR-Belarus project N183, RBFR grants 09-02-00749,
10-02-00368 and 11-02-00112.
\end{acknowledgement}


\begin{thebibliography}{99}


\bibitem {Bennett:2006fi}G.W. Bennett \textit{et al.} [Muon (g-2)
Collaboration],
Phys. Rev. D \textbf{73} (2006) 072003.

\bibitem {Jegerlehner:2011ti}F. Jegerlehner and R. Szafron,
arXiv:1101.2872 [hep-ph].

\bibitem {Davier:2010nc}M. Davier, A. Hoecker, B. Malaescu and Z. Zhang,
Eur. Phys. J. C \textbf{71} (2011) 1515.

\bibitem {Hayakawa:1995ps}M. Hayakawa, T. Kinoshita and A.I. Sanda,
Phys. Rev. Lett. \textbf{75} (1995) 790.

\bibitem {Bijnens:1995cc}J. Bijnens, E. Pallante and J. Prades,
Phys. Rev. Lett. \textbf{75} (1995) 1447.

\bibitem {Hayakawa:1997rq}M.~Hayakawa and T.~Kinoshita,
Phys.\ Rev.\ D \textbf{57} (1998) 465.

\bibitem {Knecht:2001qf}M.~Knecht and A.~Nyffeler,
Phys.\ Rev.\ D \textbf{65} (2002) 073034.

\bibitem {Melnikov:2003xd}K.~Melnikov and A.~Vainshtein,
Phys.\ Rev.\ D \textbf{70} (2004) 113006.

\bibitem {Nyffeler:2009tw}A. Nyffeler,
Phys. Rev. D \textbf{79} (2009) 073012.

\bibitem {Cappiello:2010uy}L. Cappiello, O. Cata and G. D'Ambrosio,
arXiv:1009.1161 [hep-ph].

\bibitem {Bijnens:2001cq}J.~Bijnens, E.~Pallante and J.~Prades,
Nucl.\ Phys.\ B \textbf{626} (2002) 410.

\bibitem {Bartos:2001pg}E.~Bartos, A.~Z.~Dubnickova, S.~Dubnicka, E.~A.~Kuraev
and E.~Zemlyanaya,
Nucl.\ Phys.\ B \textbf{632} (2002) 330.

\bibitem {Dorokhov:2008pw}A.~E.~Dorokhov and W.~Broniowski,
Phys.\ Rev.\ D \textbf{78} (2008) 073011.

\bibitem {Fischer:2010iz}C.~S.~Fischer, T.~Goecke and R.~Williams,
arXiv:1009.5297 [hep-ph].

\bibitem{Goecke:2010if}
T.~Goecke, C.~S.~Fischer and R.~Williams,
arXiv:1012.3886 [hep-ph].

\bibitem {Hong:2009zw}D. K. Hong and D. Kim,
Phys. Lett. B \textbf{680} (2009) 480.

\bibitem {deRafael:1993za}E. de Rafael,
Phys. Lett. B \textbf{322} (1994) 239.

\bibitem {Pallante:1994ee}E. Pallante,
Phys. Lett. B \textbf{341} (1994) 221.

\bibitem {Holdom:1993ad}B. Holdom, R. Lewis and R.R. Mendel,
Z. Phys. C \textbf{63} (1994) 71.

\bibitem {Pivovarov:2001mw}A.A. Pivovarov,
Phys. Atom. Nucl. \textbf{66} (2003) 902 [Yad. Fiz. \textbf{66} (2003) 934].

\bibitem {Dorokhov:2004ze}A.E. Dorokhov,
Phys. Rev. D \textbf{70} (2004) 094011.

\bibitem {Anikin:2000rq}I.V. Anikin, A.E. Dorokhov and L. Tomio,
Phys. Part. Nucl. \textbf{31} (2000) 509 [Fiz. Elem. Chast. Atom. Yadra
\textbf{31} (2000) 1023].

\bibitem {Dorokhov:2003kf}A.E. Dorokhov and W. Broniowski,
Eur. Phys. J. C \textbf{32} (2003) 79.

\bibitem {Scarpettini:2003fj}A.~Scarpettini, D.~Gomez Dumm and
N.~N.~Scoccola,
Phys.\ Rev.\ D \textbf{69} (2004) 114018.

\bibitem {Dorokhov:2002iu}A.~E.~Dorokhov,
JETP Lett.\ \textbf{77} (2003) 63 [Pisma Zh.\ Eksp.\ Teor.\ Fiz.\ \textbf{77}
(2003) 68].

\bibitem {Jegerlehner:2009ry}F.~Jegerlehner and A.~Nyffeler,
Phys.\ Rept.\ \textbf{477} (2009) 1.

\bibitem {Nakamura:2010zzi}K.~Nakamura \textit{et al.} [Particle Data Group],
J.\ Phys.\ G \textbf{37}, 075021 (2010).

\bibitem {Behrend:1990sr}H.~J.~Behrend \textit{et al.} [CELLO Collaboration],
Z.\ Phys.\ C \textbf{49} (1991) 401.

\bibitem {Gronberg:1997fj}J.~Gronberg \textit{et al.} [CLEO Collaboration],
Phys.\ Rev.\ D \textbf{57} (1998) 33.

\bibitem {Aubert:2009mc}B.~Aubert \textit{et al.} [BABAR Collaboration],
Phys.\ Rev.\ D \textbf{80} (2009) 052002.

\bibitem {:2011hk}P. del Amo Sanchez \textit{et al.} [BABAR Collaboration],
arXiv:1101.1142 [hep-ex].

\bibitem {Nyffeler:2009uw}A.~Nyffeler,
PoS C \textbf{D09} (2009) 080.

\bibitem{Dorokhov:2009dg}
  A.~E.~Dorokhov,
  Phys.\ Part.\ Nucl.\ Lett.\  {\bf 7 } (2010)  229-234.

\bibitem {Dorokhov:2009jd}A.E. Dorokhov,
JETP Lett. \textbf{91} (2010) 163 [Pisma Zh. Eksp. Teor. Fiz. \textbf{91}
(2010) 175].

\bibitem {Brodsky:1981rp}S.J. Brodsky and G.P. Lepage,
Phys. Rev. D \textbf{24} (1981) 1808.

\bibitem {Novikov:1983jt}V.A. Novikov, M.A. Shifman, A.I. Vainshtein, M.B.
Voloshin and V.I. Zakharov,
Nucl.\ Phys.\ B \textbf{237} (1984) 525.

\bibitem {Vainshtein:2002nv}A. Vainshtein,
Phys. Lett. B \textbf{569}, 187 (2003).

\bibitem {Dorokhov:2005pg}A.E. Dorokhov,
Eur. Phys. J. C \textbf{42} (2005) 309.

\bibitem {Dorokhov:2005hw}A.E. Dorokhov,
JETP Lett. \textbf{82} (2005) 1 [Pisma Zh. Eksp. Teor. Fiz. \textbf{82} (2005)
3].

\bibitem {Davydychev:1995mq}A.I. Davydychev and J.B. Tausk,
Phys. Rev. D \textbf{53} (1996) 7381.

\bibitem {Bijnens:2007pz}J. Bijnens and J. Prades,
Mod. Phys. Lett. A \textbf{22} (2007) 767.

\bibitem {Bijnens:1995xf}J.~Bijnens, E.~Pallante and J.~Prades,
Nucl.\ Phys.\ B \textbf{474}, 379 (1996).



\end{thebibliography}
\end{document}